# Spin-to-Orbital Conversion of Angular Momentum in Harmonic Generation Driven by Intense Circularly Polarized Beam

SHASHA LI,[1,2] XIAOMEI ZHANG,[1] WEIFENG GONG,[1,2] ZHIGANG BU,[1,*] AND BAIFEI SHEN[1,3,*]

[1]*State Key Laboratory of High Field Laser Physics, Shanghai Institute of Optics and Fine Mechanics, Chinese Academy of Sciences, Shanghai 201800, China*

[2]*University of Chinese Academy of Sciences, Beijing 100049, China*

[3]*Department of Physics, Shanghai Normal University, Shanghai, 200234, China*

*\*zhigang.bu@siom.ac.cn*

*\*bfshen@mail.shcnc.ac.cn*

**Abstract:** A new physical mechanism to achieve spin-to-orbital angular momentum conversion based on the interaction of an intense circularly polarized (CP) laser beam with a plane foil is presented and studied for the first time. It has been verified by both simulation and theoretical analysis that vortex harmonics carrying orbital angular momentum (OAM) are generated after a relativistic CP laser beam, even a Gaussian beam, impinges normally on a plane foil. The generation of this vortex harmonics is attributed to the vortex oscillation of the plasma surface driven harmonically by the vortex longitudinal electric field of the CP beam. During the process of harmonic generation, the spin angular momenta of fundamental-frequency photons are converted to OAM of harmonic photon because of the conservation of total angular momentum. In addition, if an initially vortex beam or a spiral phase plate is used, the OAM of harmonic photon can be more tunable and controllable.

The total angular momentum (TAM) of a light beam is composed of intrinsic spin angular momentum (SAM) and orbital angular momentum (OAM). The SAM is associated with the circularly polarization (CP) state. Each photon carries SAM of $\sigma = \pm 1$ ($\hbar = 1$ in this paper). The OAM is related to the spatial structure of wave front. A beam carrying OAM, named vortex beam[1], has helical-shaped wave front, and can be used as an ideal probing and manipulating tool in wide fields, such as optical manipulations[2-4], optical communications[5, 6], quantum information and computation[7], super-resolution microscopy[8], and even astrophysics[9].

In the field of optical material, some optical elements, such as q-plate[10-14], semiconductor microcavities[15], and elements using metasurface[16-19] can be used to realize spin-to-orbital angular momentum conversion by changing the direction of photon spin[20], therefore only $2\sigma$ angular momentum could be converted from SAM to OAM with these techniques. In this letter, we present a completely different mechanism to achieve the tunable and controllable conversion from spin to orbital angular momenta based on high-order harmonic generation (HHG) when an intense CP light beam interacting with a plane foil. This mechanism could provide deep insight into the nature of spin and orbital angular momenta.

According to previous results [21-23], a CP beam normally irradiating on a plane foil cannot generate harmonics. Here we demonstrate it is actually possible to generate the harmonics, because the longitudinal electric field of the CP beam induced by the finite focal size could drive the plasma surface to oscillate harmonically. Surprisingly, the generated harmonics are vortex beams. This is because the SAM of harmonic photons can only be ±1, and the extra SAMs of the fundamental-frequency light are converted to OAM due to the conservation of TAM during HHG process. In the similar way, this conversion also exists in the interaction between vortex beam and spiral phase plate (SPP) target. Therefore, the OAM of harmonic photons can be more tunable and controllable. This mechanism has some potential applications in ultra-bright, extreme ultraviolet vortex attosecond physics, optical manipulation and quantum information field.

To revealing the physical mechanism, three-dimensional particle-in-cell (PIC) simulations are performed based on EPOCH[24], where a relativistic CP Gaussian beam impinges normally on a plane foil. The driving laser beam propagates along $z$-axis, its transverse electric field is described as $\boldsymbol{E} = a_0 \exp(-r^2/w^2)\sin^2(\pi t/(2\tau))(\boldsymbol{e}_x + i\sigma \boldsymbol{e}_y)$, where $a_0 = eE_0/(m_e c\omega_0) = 12$ is the normalized dimensionless laser electric field, $m_e$ is the electron mass, $e$ is the electron charge, $\omega_0$ is the laser frequency, and c is the speed of light in vacuum. The wavelength of the laser pulse is $\lambda = 800nm$, spot size $w = 5\lambda = 4\mu m$ and the pulse duration $\tau = 5T$, where $T$ is the pulse period. The simulation box is $20\lambda(x) \times 20\lambda(y) \times 12\lambda(z)$, divided into 600×600×720 cells. The plane foil has a density of $10n_c$ and occupies the region $10\lambda < z < 12\lambda$, where $n_c = m_e \omega_0^2/(4\pi e^2)$ is the critical density.

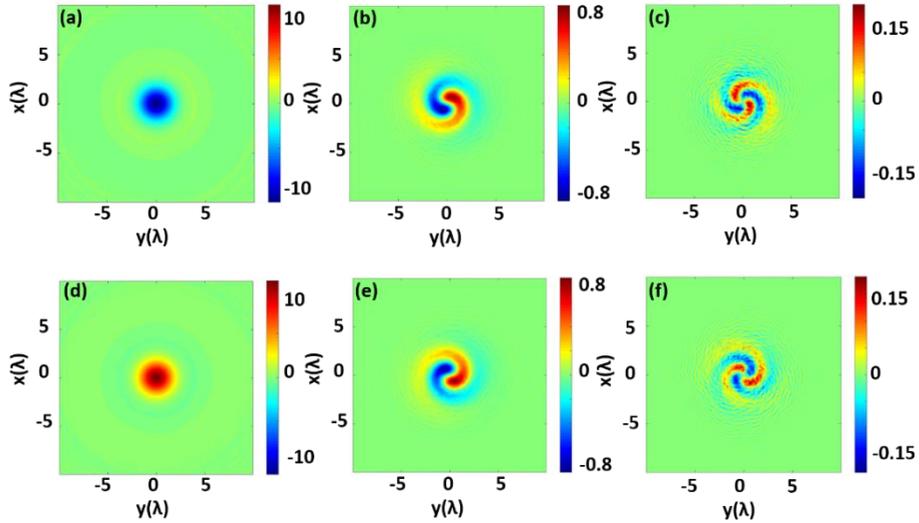

Fig. 1. Transverse electric fields $E_y$ [ $\sigma$=+1 in (a)-(c), and $\sigma$=-1 in (d)-(f)] of first [(a) and (d)], second [(b) and (e)], and third [(c) and (f)] harmonics at $z=6\lambda$ at 24$T$.

When the driving CP beam interacts with the foil, harmonics is generated in the reflected beam. The harmonics are still CP beams, which can be verified in latter theoretical analysis. Therefore, we only focus on y-component of the electric field, $E_y$. The fundamental-frequency beam in the reflected light is still Gaussian, as shown in Fig. 1(a) and 1(d). However surprisingly, Fig. 1(b)-1(c) and 1(e)-1(f) indicate the harmonics are vortex beams. In Fig.1(a)-1(c), where the driving beam is left-handed ($\sigma=+1$) and carries the TAM of $j_0=\sigma=+1$, the OAMs of first to third harmonics are $l_1 = 0$, $l_2 = +1$, $l_3 = +2$, respectively. In Fig.1(d)-1(f), where the driving beam is right-handed ($\sigma=-1$) and carries the TAM of $j_0=\sigma=-1$, the OAMs of first to third harmonics are the same but with the opposite sign. These results are easily understood from the view of the conservation of TAM. In the HHG process, multiple fundamental-frequency photons of the driving beam are absorbed by the oscillating plasma surface and converted into a single harmonic photon. So, the TAM carried by the photon of $n_{th}$ harmonic should be $n$ times that of the driving beam, i.e. $j_n = n\sigma$. Since the SAM of a photon can only be ±1, the extra SAMs would be converted to OAM to ensure the conservation of TAM, which leads to $l_n = (n-1)\sigma$.

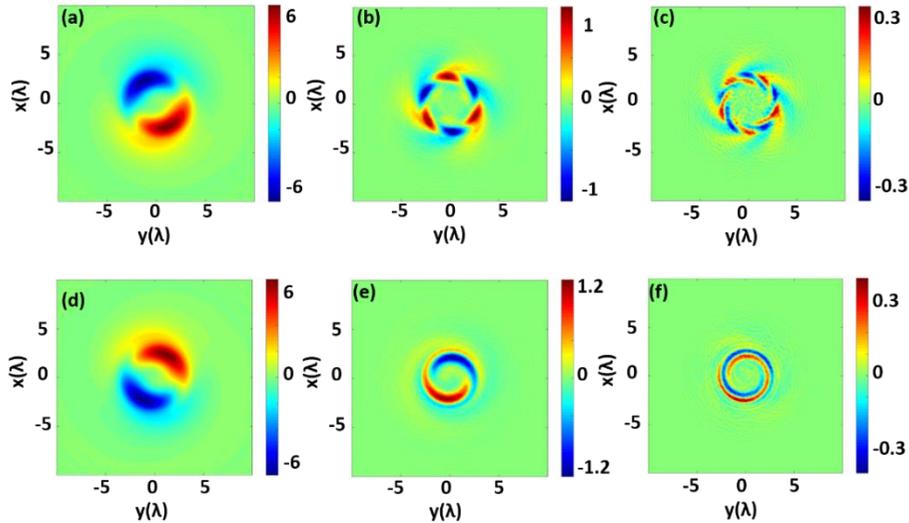

Fig. 2. The harmonics generated by the left-handed CP LG beam. Transverse electric fields $E_y$ [ $l$=+1 in (a)-(c), and $l=-1$ in (d)-(f)] of the fist [(a) and (d)], second [(b) and (e)], and third [(c) and (f)] harmonics of the reflected beams at $z=6\lambda$ at 24T.

The conversion of SAM to OAM not only exists in the HHG driven by a CP Gaussian beam, but also happens for vortex beams. Fig. 2 illustrates two groups of harmonics generated by left-handed CP Laguerre-Gaussian (LG) beams with different OAMs, where the normalized laser electric field is $a_0 = 20$ and other parameters are the same with the Gaussian beam above (to keep the same peak power). In the first case, the driving beam carries the OAM of $l_0 = +1$, and the OAMs of first to third harmonics are $l_1 = +1$, $l_2 = +3$ and $l_3 = +5$, as shown in Fig. 2(a)-2(c). In the second case shown in Fig. 2(d)-2(f), the OAM of the driving beam is $l_0 = -1$, which leads the OAMs of all the harmonics to be $|l_n| = 1$. These results can be explained as follows: The driving beam carries the TAM of $j_0 = \sigma + l_0$. According to the TAM conservation, the TAM of $n_{th}$ harmonics is determined by $j_n = nj_0 = n(\sigma + l_0)$. Since the SAM of $n_{th}$ harmonic photon is still $\sigma = +1$, then the OAM of $n_{th}$ harmonic photon is given by $l_n = j_n - \sigma = (n-1)\sigma + nl_0$. In the first case, the OAM of the driving beam is $l_0 = +1$, so we get $l_n = 2n-1$; while driving beam in the second case carries the OAM of $l_0 = -1$, which leads to a constant OAM of $l_n = -1$ for each order of harmonics.

From the above simulation results, it can be concluded that since the SAM of a CP photon can only be +1 or -1, the OAM of $n_{th}$ harmonic photon should be $l_n = (n-1)\sigma + nl_0$ to ensure the conservation of TAM. That means the extra SAMs of the fundamental-frequency photons have been converted to OAMs of the harmonic photons.

If a vortex beam impinges normally on a SPP with a step height $h = l_s\lambda$, additional OAM of $l_s$ is produced [25]. Then the TAM of the fundamental-frequency beam becomes $j_0 = \sigma + l_0 + l_s$. Based on our analysis, the OAM of nth harmonic photon would be $l_n = (n-1)\sigma + n(l_0 + l_s)$. In this way, the OAM of $n_{th}$ harmonic photon can be tuned and controlled by changing these three parameters. And a low order harmonic with a high OAM is achievable. Table 1 shows the simulation results under different conditions. The plane foil and Gaussian beam corresponding to $l_s$=0 and $l_0$=0, respectively. Simulations indicate that the OAM of $n_{th}$ harmonic photon follows the rules: $l_n = (n-1)\sigma + n(l_0 + l_s)$. These results demonstrate the spin-to-orbital angular momentum conversion occurs during the HHG driven by a CP beam under different conditions.

**Table 1. The OAM of harmonic photons under different conditions.**

| SPP | SAM and OAM of Incident Photon | | OAM of Harmonic Photons | | |
|---|---|---|---|---|---|
| $l_s$ | $\sigma$ | $l_0$ | $l_1$ | $l_2$ | $l_3$ |
| 0 | +1 | 0 | 0 | +1 | +2 |
| 0 | -1 | 0 | 0 | -1 | -2 |
| 0 | +1 | -1 | -1 | -1 | -1 |
| 0 | +1 | +1 | +1 | +3 | +5 |
| +1 | -1 | +1 | +2 | +3 | +4 |
| +1 | +1 | +1 | +2 | +5 | +8 |
| +1 | -1 | -2 | -1 | -3 | -5 |
| +1 | +1 | -2 | -1 | -1 | -1 |

Besides, it should be noted that the harmonic is stronger when we use a vortex driving beam or SPP target, because the phase plane of LG laser and the SPP surface are helical, which means the beam is incident obliquely at a small angle and the harmonic generation efficiency for CP driving beam is increased.

In physics, the spin-to-orbital angular momentum conversion is required by the conservation of TAM. The details of how the harmonics are generated are not important for above phenomenon. Even for HHG due to laser atom interaction, same results should be obtained. But here to understand the mechanism how the SAM is converted to OAM in our condition, let us analyze the HHG induced by the CP beam in theory firstly. In Coulomb gauge, the vector potential of the driving CP beam can be expressed in a standing wave form due to the reflection on the foil,

$$A(t,x,y,z) = A_0 \begin{pmatrix} u(r)\sin(k_0 z)\sin(\omega_0 t) \\ -\sigma u(r)\sin(k_0 z)\cos(\omega_0 t) \\ k_0^{-1}\partial_r u(r)\cos(k_0 z)\sin(\omega_0 t - \sigma\theta) \end{pmatrix}, \quad (1)$$

where $A_0$ is the amplitude, $r = \sqrt{x^2 + y^2}$, $u(r)$ is the transverse profile and only determined by $r$ due to the cylindrical symmetry, $\theta = \arctan(y/x)$ is the azimuth angle, $\sigma = \pm 1$ is the SAM, $k_0$ and $\omega_0$ are the wave vector and frequency of the beam. Here $u(r) = \exp(-r^2/w^2)$ describes the transverse profile of Gaussian beam. Then the electric field is obtained as $E(t,x,y,z) = -(1/c)\partial_t A$,

$$E(t,x,y,z) = -A_0 \begin{pmatrix} k_0 u(r)\sin(k_0 z)\cos(\omega_0 t) \\ \sigma k_0 u(r)\sin(k_0 z)\sin(\omega_0 t) \\ \partial_r u(r)\cos(k_0 z)\cos(\omega_0 t - \sigma\theta) \end{pmatrix}. \quad (2)$$

It is well known that the HHG results from the collective oscillating motion of electrons. In the existed harmonic generation theory [21-23], harmonics are radiated by the oscillating plasma surface driven by the pondermotive force of the laser beam. For a linearly polarized laser, the ponderomotive force oscillates at twice the frequency of driving laser beam, thus only odd harmonics are generated when the driving laser impinges normally on the foil. However, for a CP laser, the ponderomotive force is constant, and no harmonics are generated at all. Here, the mechanism of the harmonic generation is completely different, Eq. (2) indicates a longitudinal electric field, $E_z$, is induced because of the finite transverse profile of the CP laser beam. This longitudinal electric field contains a rapid oscillation factor with the laser frequency, and drives the plasma surface to oscillate at the same frequency. Consequently, both odd and even harmonics can be generated. In addition, $E_z$ has a vortex phase $\exp(i\sigma\theta)$ similar to an LG beam with a topological charge of 1 as shown in Fig. 3, which plays an important role in the angular momentum conversion in HHG process. We notice that $E_z$ is proportional to $\partial_r u(r)$. If the spot size is much large than the wavelength, we get $\partial_r u(r) \ll k_0 u(r)$, and the conversion efficiency of harmonics is low. In the limit of plane wave, $E_z \to 0$, there are no harmonics generated. However, when the spot size is small and compared with the laser wavelength, we get $\partial_r u(r) \leq k_0 u(r)$, that means $E_z$ could be compared with the transverse electric field of the driving beam, and the conversion efficiency is increased evidently. In addition, $E_z$ can also induce longitudinal ponderomotive force. But this ponderomotive force is much smaller than $E_z$, and can be ignored.

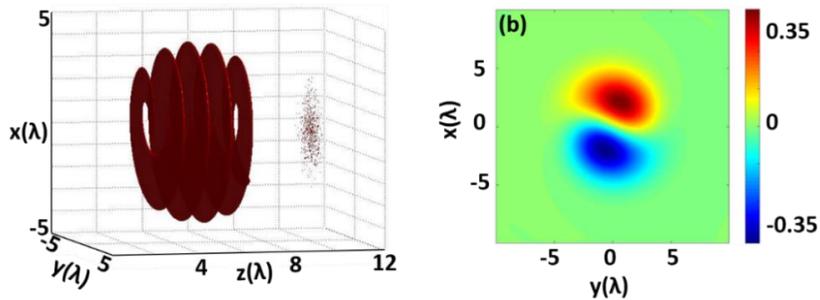

Fig. 3. The longitudinal electric field, $E_z$, of the left-handed CP Gaussian beam. (a) The isosurface of $E_z$. (b) Transverse section of $E_z$ at $z=6\lambda$ at $14T$.

Next, we focus on the physical mechanism of angular momentum conversion in HHG process. The plasma surface is driven harmonically by the longitudinal electric field of the CP beam, and its motion can be described by $Z(t_{ret})$ [22], where $t_{ret} = t - Z(t_{ret})/c$ is the retarded time at the observation point $z_{obs}$. Under the lowest order approximation we obtain

$$Z(t_{ret}) = -Z_s \sin(\omega_0 t - \sigma\theta), \qquad (3)$$

the polarization of the harmonic is the same as the driving light beam, and the transverse components of the reflected electric field is given by

$$\mathcal{E}_\perp(z_{obs},t) \sim \mathcal{E}_0 a_\perp(Z(t_{ret}),t_{ret}) \sim \begin{pmatrix} \sin(\omega_0 t + k_0 Z_s \sin(\omega_0 t - \sigma\theta)) \\ -\sigma\cos(\omega_0 t + k_0 Z_s \sin(\omega_0 t - \sigma\theta)) \end{pmatrix}, \qquad (4)$$

where $Z_s$ is the amplitude of the surface motion, $\mathbf{a} = e\mathbf{A}/(mc)$ is the normalized vector potential of the CP beam, and the transverse structure of the reflected field is neglected. Fourier expansion of Eq. (4) gives two angular momentum modes of harmonic components, $\mathcal{E}_\perp(z_{obs},t) = \mathcal{E}_\perp^{(\sigma)}(z_{obs},t) + \mathcal{E}_\perp^{(-\sigma)}(z_{obs},t)$,

$$\mathcal{E}_\perp^{(\sigma)}(z_{obs},t) \sim \sum_{n=1}^{\infty} J_{n-1}(kZ_s) \begin{pmatrix} \sin[n\omega_0 t - (n-1)\sigma\theta] \\ -\sigma\cos[n\omega_0 t - (n-1)\sigma\theta] \end{pmatrix} = \sum_{n=1}^{\infty} \varepsilon_\perp^{(\sigma)}(n\omega_0), \quad (5a)$$

$$\mathcal{E}_\perp^{(-\sigma)}(z_{obs},t) \sim \sum_{n=1}^{\infty} (-1)^n J_{n+1}(kZ_s) \begin{pmatrix} \sin[n\omega_0 t - (n+1)\sigma\theta] \\ \sigma\cos[n\omega_0 t - (n+1)\sigma\theta] \end{pmatrix} = \sum_{n=1}^{\infty} \varepsilon_\perp^{(-\sigma)}(n\omega_0), \quad (5b)$$

where $J_\nu(s)$ denotes Bessel function of first kind. Eq. 5(a) and 5(b) represent two different angular momentum modes of harmonics with OAM of $l_n = (n \pm 1)\sigma$. These two modes are closely related to the SAM of the generated harmonics, which can be explained by TAM conservation. Since the driving beam carries no OAM, its TAM is only determined by SAM, $j_0 = \sigma$. During the HHG process, the $n_{th}$ harmonic photon carries the TAM of $j_n = n\sigma$. If the harmonic photon carries the same SAM as the incident photons in driving beam, the extra SAMs would be converted to OAM due to the TAM conservation, which leads to $l_n = (n-1)\sigma$ as formulated by mode $\varepsilon_\perp^{(\sigma)}(n\omega_0)$ in Eq. (5a). However, there is another case. If the SAM of the harmonic photon flips to $-\sigma$, the OAM must be $l_n = (n+1)\sigma$, which is expressed by mode $\varepsilon_\perp^{(-\sigma)}(n\omega_0)$ in Eq. (5b).

Since the amplitude of the plasma surface motion is small, we have $k_0 Z_s \ll 1$. Then the Bessel function in Eq. (5) can be expressed by the limiting form: $J_n(k_0 Z_s) \sim (k_0 Z_s/2)^n / \Gamma(n+1)$, where $\Gamma(s)$ is the Gamma function. Thus, the electric field of $n_{\text{th}}$ harmonic mode $\varepsilon_\perp^{(-\sigma)}(n\omega_0)$ is much lower than mode $\varepsilon_\perp^{(\sigma)}(n\omega_0)$, with the decay factor $\left|\varepsilon_\perp^{(-\sigma)}(n\omega_0)/\varepsilon_\perp^{(\sigma)}(n\omega_0)\right| \sim (k_0 Z_s)^2/(4n(n+1)) \ll 1$. Therefore, the harmonic with mode $\varepsilon_\perp^{(-\sigma)}(n\omega_0)$ can be neglected.

When the driving light beam is the CP LG beam, the vector potential is written as

$$\boldsymbol{A}_{LG} = A_0 \begin{pmatrix} u_{LG}(r)\sin(k_0 z)\sin(\omega_0 t - l_0\theta) \\ -\sigma u_{LG}(r)\sin(k_0 z)\cos(\omega_0 t - l_0\theta) \\ k_0^{-1}\left(\partial_r u_{LG}(r) - (\sigma l_0/r) u_{LG}(r)\right)\cos(k_0 z)\sin(\omega_0 t - j_0\theta) \end{pmatrix} \quad (6)$$

in Coulomb gauge, where $j_0 = \sigma + l_0$ is TAM, $u_{LG}(r) = \left(\sqrt{2}r/w\right)^{|l_0|}\exp(-r^2/w^2) L_p^{|l_0|}(2r^2/w^2)$ and $L_p^l(s)$ is associated Laguerre polynomial. Based on the similar theoretical analysis discussed above, the two angular momentum modes of harmonic components are obtained, with the results

$$\mathcal{E}_{LG\perp}^{(\sigma)}(z_{obs},t) \sim \sum_{n=1}^{\infty} J_{n-1}(k_0 Z_s) \begin{pmatrix} \sin(n\omega_0 t - l_n^{(\sigma)}\theta) \\ -\sigma\cos(n\omega_0 t - l_n^{(\sigma)}\theta) \end{pmatrix}, \quad (7a)$$

$$\mathcal{E}_{LG\perp}^{(-\sigma)}(z_{obs},t) \sim \sum_{n=1}^{\infty}(-1)^n J_{n+1}(k_0 Z_s)\begin{pmatrix}\sin(n\omega_0 t - l_n^{(-\sigma)}\theta) \\ \sigma\cos(n\omega_0 t - l_n^{(-\sigma)}\theta)\end{pmatrix}, \quad (7b)$$

where $l_n^{(\pm\sigma)} = nl_0 + (n\mp 1)\sigma$ are the OAM of $n_{\text{th}}$ harmonic photons for modes $\mathcal{E}_{LG\perp}^{(\sigma)}$ and $\mathcal{E}_{LG\perp}^{(-\sigma)}$, respectivly, and they obey TAM conservation. However, same as the situation of CP Gaussian beam, the harmonic with mode $\mathcal{E}_{LG\perp}^{(-\sigma)}$ can be neglected. This is just the reason why the SAM of the harmonic is always the same as the driving light beam in our simulations. In addition, for the LG beam, there is an additional contribution to the longitudinal electric field: $\sim (\sigma l_0/r)u_{LG}(r)$, which means the conversion efficiency of the harmonics can be further increased compared with that in the Gaussian beam case.

In conclusion, we present a new mechanism to achieve the spin-to-orbital angular momentum conversion based on HHG process when an intense CP beam interacting with a plane foil. In the HHG, since the SAM of a photon can only be $\pm 1$, the extra SAMs of fundamental-frequency photons would be converted to OAM due to the conservation of TAM. This mechanism has been verified by the simulation and theoretical analysis, it provides a deep insight into spin-to-orbital angular momentum conversion. The presented angular momentum conversion has some potential applications in ultra-bright, extreme ultraviolet vortex attosecond physics, optical manipulation and quantum information field. This work opens a new window in study of the spin-orbital interaction in optics.

We are grateful to Jingwei Wang for many helpful discussions. This work is supported by Ministry of Science and Technology of the People's Republic of China (2018YFA0404803 and 2016YFA0401102), the National Natural Science Foundation of China (11674339) and Strategic Priority Research

Program of the Chinese Academy of Sciences (XDB16). We also thank for the support of Innovation Program of Shanghai Municipal Education Commission and Shanghai Supercomputer Center.[1] L. Allen, M.W. Beijersbergen, R.J.C. Spreeuw, and J.P. Woerdman, Phys. Rev. A **45**, 8185 (1992).

[2] T. Kuga, Y. Torii, N. Shiokawa, T. Hirano, Y. Shimizu, and H. Sasada, Phys. Rev. Lett. **78**, 4713 (1997).

[3] M. Padgett, Proc. R. Soc. A. **470**(2172) (2014).

[4] D.G. Grier, Nature, **424**, 810 (2003).

[5] J. Wang, Photon. Res. **5**, 19 (2016).

[6] J. Wang, J-Y. Yang, I.M. Fazal, N. Ahmed, Y. Yan, H. Huang, Y. Ren, Y. Yue, S. Dolinar, M. Tur, and A. E. Willner, Nature photon. **6,** 488 (2012).

[7] A. Mair, A. Vaziri, G. Weihs, and A. Zeilinger, Nature **412,** 313 (2001).

[8] K. Ladavac, and D.G. Grier, Opt. Express **12,** 1144 (2004).

[9] F. Tamburini, B. Thidé, G. Molina-Terriza, and G. Anzolin, Nature Phys. **7,** 195 (2011).

[10] L. Marrucci, C. Manzo, and D. Paparo, Phys. Rev. Lett. **96,** 163905 (2006).

[11] B. Piccirillo, V. D'Ambrosio, S. Slussarenko, L. Marrucci, and E. Santamato, Appl. Phys. Lett. **97**, 241104 (2010).

[12] M. Rafayelyan, and E. Brasselet, Phys. Rev. Lett. **120,** 213903 (2018).

[13] S. Slussarenko, A. Murauski, T. Du, V. Chigrinov, L. Marrucci, and E. Santamato, Opt. Express **19,** 4085 (2011).

[14] E. Nagali, F. Sciarrino, F. De Martini, L. Marrucci, B. Piccirillo, E. Karimi, and E. Santamato, Phys. Rev. Lett. **103,** 013601 (2009).

[15] F. Manni, K.G. Lagoudakis, T.K. Paraïso, R. Cerna, Y. Léger, T.C.H. Liew, I.A. Shelykh, A.V. Kavokin, F. Morier-Genoud, and B. Deveaud-Plédran, Phys. Rev. B
12